\documentclass[a4paper]{article}
\title{Supplementation of reducible constraints 
and the Green-Schwarz superstring.}
\author{A.A. Deriglazov\thanks{alexei@fisica.ufjf.br ~ On leave of
absence from Dept. Math. Phys., Tomsk Polytechnical University,
Tomsk, Russia}}
\date{Instituto de F\'\i sica, Universidade Federal de Juiz de Fora,\\
MG, Brasil.}
\begin{document}
\maketitle
\large
\begin{abstract}
We apply the supplementation trick [26] to the Green-Schwarz superstring. 
For type IIB theory both first and second class
constraints are covariantly separated and then arranged into irreducible 
sets in the initial formulation.
For $N=1$ Green-Schwarz superstring we propose a modified action 
which is equivalent to the initial one.    
Fermionic first and second class constraints are covariantly 
separated, the first class constraints (1CC) turn out to be irreducible. 
We discuss also equations of motion in the covariant gauge for 
$\kappa$-symmetry.  
For type IIA theory the same modification leads to formulation with 
irreducible second class constraints.
\end{abstract}

{\bf PAC codes:} 0460D, 1130C, 1125 

\noindent
\section{Introduction}
Manifestly super Poincare invariant formulation of branes implies 
appearance of mixed first and second class fermionic constraints 
in the Hamiltonian formalism [1-10]. Typically, 
first and second class constraints (2CC) are treated in a rather 
different way in quantum theory\footnote{Quantization scheme for mixed 
constraints was developed in [11]. Application of this scheme to concrete 
models may conflict with manifest Poincare covariance [12].}. In 
particular, to construct formal expression for the covariant path 
integral one needs to have splitted and irreducible constraints [13, 14]. 
So, it is necessary at first to split them, 
which can be achieved 
by using of covariant projectors of one or other kind [15-17, 12]. 
Details depend on the model under consideration. For example, for 
CBS superparticle [18, 1] one introduces two auxilliary vector 
variables in addition to the initial superspace coordinates [15]. 
For the Green-Schwarz (GS) superstring the projectors can be constructed 
in terms of the initial variables only [16]. After that the problem 
reduces to quantization of the covariantly separated but 
{\em infinitely} reducible constraints. Despite a lot of efforts 
(see [15-23] and references therein) this problem has no fully 
satisfactory solution up to date\footnote{Infinitely reducible 1CC 
imply infinite tower of ``ghosts for ghosts'' variables [22]. For 
reducible 2CC the problem is  
that the covariant Dirac bracket obeys the Jacobi identity 
on the second class constraints surface only [12].}. A revival of interest 
to the problem is due to recent work [24] where it was shown that 
scattering amplitudes for superstring can be constructed in a manifestly 
covariant form, as well as due to progress in the light-cone 
quantization of superstring on $AdS_5\times S^5$ background [25, 24].

One possibility to avoid the problem of quantization of infinitely 
reducible constraints is the supplementation trick which was formulated 
in the Hamiltonian framework in [26]. The basic idea is to introduce 
finite number of an additional fermionic variables subject to their 
own reducible 
constraints (the constraints are chosen in such a way that the additional 
sector do not contains physical degrees of freedom). Then the original 
constraints can be combined with one from the additional sector into 
irreducible set. For the resulting 1CC one imposes a covariant 
and irreducible 
gauge. It implies, in particular, a possibility to construct correct 
Dirac bracket for the theory. 

To apply the recipe for concrete model one needs to find a modified 
Lagrangian which reproduces the desired irreducible constraints. 
Some examples were considered [26, 27], in particular, the modified 
$N=1$ GS superstring  action was proposed in [28]. But it was pointed 
in [29]  that the action is not equivalent to the initial one.

In this work we present modified action which is equivalent to $N=1$ 
GS superstring and which allows one to realize the supplementation 
trick. Then we analyse type II GS superstring.  

The work is organized as follows. To fix our notations, we review main 
steps of the supplementation scheme in Sec. 2. 
In Sec. 3 modified formulation
of $N=1$ GS superstring is presented and proved to be
equivalent to the initial one. First and second class constraints are
covariantly separated, 1CC form irreducible set. We discuss also 
equations of motion
and their solution in the covariant gauge for $\kappa$-symmetry. 
It is shown how the usual Fock space picture can be obtained in this 
gauge. Type IIB theory is considered in Sec. 4. Here one 
has two copies of the fermionic constraints (which correspond
to two $\theta^A, ~ A=1,2$) with the same chirality. It allows one to
consider their Poincare covariant combinations. In this case 
{\em{both}} first
and second class constraints can be arranged into covariant sets in the
initial formulation. For type IIA theory (Sec. 5) the two copies of
constraints have an opposite chirality and can not be combined in the
initial formulation. Repeating the same steps as in $N=1$ case, one
finds that all the second class constraints as well as one chiral sector
of the first class constraints can be combined into irreducible sets.
Other sector with first class constraints remains reducible. 
Some technical details are omitted and can be find 
in [34].

\section{Supplementation of the reducible constraints.} 

It will be convenient to work in 16-component formalism of the Lorentz
group $SO(1, 9)$,
then $\theta^\alpha$, $\psi_\alpha$, $\alpha=1,\dots,16$, are
Majorana--Weyl spinors of opposite chirality. Real, symmetric
$16\times16$ $\Gamma$-matrices ${\Gamma^\mu}_{\alpha\beta}$,
$\tilde\Gamma^{\mu\alpha\beta}$ obey the algebra
$\Gamma^\mu\tilde\Gamma^\nu+\Gamma^\nu\tilde\Gamma^\mu=-2\eta^{\mu\nu},
~ \eta^{\mu\nu}=(+,-, \ldots ,-)$.
Momenta conjugate to configuration space variables $q^i$
are denoted as $p_{qi}$.

Let us consider a dynamical system with fermionic pairs $(\theta^\alpha,
p_{\theta\alpha})$ being presented among the phase space variables $z^A$.
Typical situation for the models under consideration is that the following
constraints
\begin{eqnarray}\label{1}
L_\alpha\equiv p_{\theta\alpha}- iB_\mu{\Gamma^\mu}_{\alpha\beta}
\theta^\beta\approx 0,
\end{eqnarray}
\begin{eqnarray}\label{3}
D^\mu D_\mu\approx0, \qquad 
\Lambda^\mu\Lambda_\mu\approx0, 
\end{eqnarray}
are presented among others.
Here, the $B^\mu(z), ~ D^\mu(z), ~ \Lambda^\mu(z)$ 
are some functions of phase variables
$z$, so that $D^2\approx0$, ~ $\Lambda^2\approx 0$ are first class 
constraints\footnote{In some 
cases (for example, for the superparticle) the quantity $\Lambda^\mu$ 
is absent in the initial formulation. Then one needs to introduce an 
additional vector variable [26].}. It is supposed also 
$(D\Lambda)\ne 0$ which is true for the models considered below. 
Poisson bracket of the fermionic constraints is
\begin{eqnarray}\label{2}
\{L_\alpha,L_\beta\}=2iD_\mu{\Gamma^\mu}_{\alpha\beta}.
\end{eqnarray}
The system $L_\alpha\approx0$ is mixture of first and second class
constraints, as it will be proved momentarily. 

Below the following two simple facts will be used 
systematically (see [34] for the proof). \\
1). ~ Let $\Psi^\alpha=0$ are 16 equations. Then: a) The system 
\begin{eqnarray}\label{4}
D^\mu\Gamma^\mu\Psi=0,
\end{eqnarray}
\begin{eqnarray}\label{5}
\Lambda^\mu\Gamma^\mu\Psi=0,
\end{eqnarray}
is equivalent to $\Psi^\alpha=0$. \\
b) ~ Let $\Psi^\alpha=0$ represent 16 independent equations. Then 
Eq.(\ref{4}) contains 8 independent equations. In $SO(8)$ notations 
they mean that $8_s$ part $\bar\Psi_{\dot a}$ of $\Psi^\alpha$ can be 
presented through $8_c$ part $\Psi_a$ (or vice-versa). The same is 
true for Eq.(\ref{5}). \\
2). ~ Let $\Psi^\alpha=0, ~ \Phi^\alpha=0$ are $16+16$ independent 
equations. Consider the system 
\begin{eqnarray}\label{6}
D^\mu\Gamma^\mu\Psi=0, \qquad
\Lambda^\mu\Gamma^\mu\Phi=0.
\end{eqnarray}
Then: a) Eq.(\ref{6}) contains 16 independent equations according 
to 1). \\
b) ~ The equations 
\begin{eqnarray}\label{7}
D^\mu\Gamma^\mu\Psi+\Lambda^\mu\Gamma^\mu\Phi=0,
\end{eqnarray}
are equivalent to the system (\ref{6}). Thus the system (\ref{7}) 
consist of 16 independent equations (i.e. it is irreducible). 

Supplementation scheme
for the mixed constraints consist of the following steps.

{\em A). Manifestly covariant separation of the constraints.} \\
By virtue of the statement 1), the system (\ref{1}) can be rewritten 
in the equivalent form\footnote{To split the constraints one can
also use true projectors instead of the matrices $D_\mu\Gamma^\mu, ~
\Lambda_\mu\Gamma^\mu$. It allows one to avoid possible
``second class patalogy'' [30] in the 1CC algebra [16].}
\begin{eqnarray}\label{8}
L^{(1)\alpha}\equiv D_\mu\tilde\Gamma^{\mu\alpha\beta}L_\beta
\approx 0,
\end{eqnarray}
\begin{eqnarray}\label{9}
L^{(2)\alpha}\equiv \Lambda_\mu\tilde\Gamma^{\mu\alpha\beta}L_\beta
\approx 0.
\end{eqnarray}
Eq.(\ref{8}) contains 8 independent constraints  
which turn out to be 1CC as a consequence of Eqs.(\ref{3}), 
(\ref{2}). Similarly, Eq.(\ref{9}) contains 8 independent constraints.
One can rewrite them in $SO(8)$ notations and verify that they form 
2CC system. 

{\em B). Auxiliary sector subject to reducible constraints.} \\ 
Let us introduce a pair of spinors
$(\eta^\alpha,p_{\eta\alpha})$ subject to the constraints
\begin{eqnarray}\label{10}
p_{\eta\alpha}\approx0, \qquad T_\alpha\equiv \Lambda_\mu
{\Gamma^\mu}_{\alpha\beta}\eta^\beta\approx0.
\end{eqnarray}
These equations contain 8 independent 1CC among $p^{(1)}_\eta\equiv
\Lambda_\mu\tilde\Gamma^\mu p_\eta\approx0$ and 8+8 independent 2CC
among $p^{(2)}_\eta\equiv D_\mu\tilde\Gamma^\mu p_\eta\approx0$,
$\Lambda_\mu\Gamma^\mu\eta\approx0$. Note that the covariant gauge
$D_\mu\Gamma^\mu\eta=0$ may be imposed. After that the complete system
(constraints + gauges) is equivalent to $p_\eta\approx0$,
$\eta\approx0$.

{\em C). Supplementation up to irreducible constraints.} \\ 
Now part of the constraints can be combined into covariantly separated
and irreducible sets. According to the statement 2), the system 
(\ref{8})-(\ref{10}) is equivalent to
\begin{eqnarray}\label{11}
\Phi^{(1)\alpha)}\equiv L^{(1)\alpha}+p^{(1)\alpha}_\eta=D_\mu
\tilde\Gamma^{\mu\alpha\beta}L_\beta+\Lambda_\mu
\tilde\Gamma^{\mu\alpha\beta} p_{\eta\beta}\approx 0,
\end{eqnarray} 
\begin{eqnarray}\label{12}
\Phi^{(2)\alpha}\equiv L^{(2)\alpha}+p^{(2)\alpha}_\eta=\Lambda_\mu
\tilde\Gamma^{\mu\alpha\beta}L_\beta+D_\mu
\tilde\Gamma^{\mu\alpha\beta} p_{\eta\beta}\approx0, 
\end{eqnarray}
\begin{eqnarray}\label{13}
T_\alpha\equiv \Lambda_\mu{\Gamma^\mu}_{\alpha\beta}\eta^\beta
\approx0,
\end{eqnarray}
where $\Phi^{(1)\alpha}\approx0$ 
($\Phi^{(2)\alpha}\approx0$) are 16 irreducible 1CC
(2CC) and $T_\alpha\approx0$ contains 8 linearly independent 2CC.
In the result first class constraints of the extended formulation form 
irreducible set (\ref{11}). As it was mentioned above, the type IIB 
superstring presents an example of more attractive
situation as compare to the general case (\ref{11})-(\ref{13}). Due to
special structure of the theory the 2CC can also be combined into
irreducible set.

\section{$N=1$ Green-Schwarz superstring with irreducible first 
class constraints.}

Consider GS superstring action with $N=1$ space-time supersymmetry
\begin{eqnarray}\label{34}
S=-\frac{T}{2}\int d^2\sigma
\left[\frac{1}{\sqrt{-g}}g^{ab}
\Pi_a^\mu \Pi_b^\mu+2i\varepsilon^{ab}\partial_ax^\mu
\theta\Gamma^\mu\partial_b\theta\right],
\end{eqnarray}
where $\sqrt{-g}=\sqrt{-\det{g^{ab}}}, ~ \Pi^\mu_a\equiv
\partial_ax^\mu-i\theta\Gamma^\mu\partial_a\theta, ~
\varepsilon^{01}=-1$. Let us denote
\begin{eqnarray}\label{35}
B^\mu\equiv p^\mu+T\Pi_1^\mu, \qquad
\hat p^\mu\equiv p^\mu-iT\theta\Gamma^\mu\partial_1\theta, \cr
D^\mu\equiv\hat p^\mu+T\Pi_1^\mu, \qquad
\Lambda^\mu\equiv\hat p^\mu-T\Pi_1^\mu.
\end{eqnarray}
Then constraints under the interest for Eq.(\ref{34}) are those of 
Eqs.(\ref{1}), (\ref{3}).
From Eqs.(\ref{35}), (\ref{3}) and from the standard requirement that
the induced metric is non degenerated it follows 
$D\Lambda=\hat p^2-T\Pi_1^2\ne 0$. 
Using the quantities $D^\mu, ~ \Lambda^\mu$, the fermionic 
constraints $L_\alpha$ can be decomposed now on first and second class 
subsets as in Eq.(\ref{8}), (\ref{9}). To proceed further, one needs to 
find a modification which will lead to Eq.(\ref{10}).

The modified action which we propose below acquires more elegant form 
in ADM representation for the world-sheet metric: 
$g^{00}=\gamma^{-1}N^{-2}, ~ 
g^{01}=\gamma^{-1}N^{-2}N_1, ~
g^{11}=\gamma^{-1}N^{-2}(N_1^2-N^2)$. 
The GS action (\ref{34}) acquires then the
following form:
\begin{eqnarray}\label{42}
S=-\frac{T}{2}\int d^2\sigma
\left[\frac{1}{N}
\Pi_+^\mu \Pi_-^\mu+2i\varepsilon^{ab}\partial_ax^\mu
\theta\Gamma^\mu\partial_b\theta\right],
\end{eqnarray}
where it was denoted 
$\Pi^\mu_{\pm}\equiv\Pi_0^\mu+N_{\pm}\Pi_1^\mu, ~ 
N_{\pm}\equiv N_1\pm N$.
The world-sheet reparametrisations in this representation look as 
$\delta\sigma^a=\xi^a, ~  \delta N_{\pm}=\partial_0\xi^1+
(\partial_1\xi^1-
\partial_0\xi^0)N_{\pm}-\partial_1\xi^0N_{\pm}^2$. 
Modified action to be examined is
\begin{eqnarray}\label{44}
S=-\frac{T}{2}\int d^2\sigma
\left[\frac{1}{N}
\Pi_+^\mu (\Pi_-^\mu+i\eta\Gamma^\mu\chi)+
2i\varepsilon^{ab}\partial_ax^\mu
\theta\Gamma^\mu\partial_b\theta-\frac{1}{4N}
(\eta\Gamma^\mu\chi)^2\right],
\end{eqnarray}
where two additional Majorana-Weyl fermions
$\eta^\alpha(\tau, \sigma), ~ \chi^\alpha(\tau, \sigma)$ were introduced.
Our aim now will be to show canonical equivalence of this action and
the initial one. Then the additional sector will be used to arrange
1CC of the theory into irreducible set.
Direct application of the Dirac algorithm gives us the Hamiltonian
\begin{eqnarray}\label{45}
H=\int d\sigma\left[-\frac N2\left(\frac {1}{T} \hat p^2+
T\Pi_1^2\right)-N_1(\hat p\Pi_1)-\right. \cr
\left.\frac i2 (\hat p^\mu-T\Pi_1^\mu)\eta\Gamma^\mu\chi+
\lambda_N p_N+\lambda_{N1}p_{N1}+\lambda_\eta p_\eta+
\lambda_\chi p_\chi+L\lambda_\theta\right],
\end{eqnarray}
where $\lambda_q$ are the Lagrangian multipliers for the corresponding
primary constraints. After determining of secondary constraints, complete
constraint system can be presented as (in the notations
(\ref{35}))
\begin{eqnarray}\label{46}
p_N=0, \qquad p_{N1}=0,
\end{eqnarray}
\begin{eqnarray}\label{47}
D^2-4TL\partial_1\theta=0, \quad
\Lambda^2-4T\partial_1\eta p_\eta-4T\partial_1\chi p_\chi=0,
\end{eqnarray}
\begin{eqnarray}\label{48}
L_\alpha\equiv p_{\theta\alpha}- iB_\mu{\Gamma^\mu}_{\alpha\beta}
\theta^\beta=0,
\end{eqnarray}
\begin{eqnarray}\label{49}
\Lambda^\mu(\Gamma^\mu\eta)_\alpha=0, \qquad
p_{\eta\alpha}=0,
\end{eqnarray}
\begin{eqnarray}\label{50}
\Lambda^\mu(\Gamma^\mu\chi)_\alpha=0, \qquad
p_{\chi\alpha}=0.
\end{eqnarray}
Note that the desired constraints (\ref{10}) appear in duplicate form
(\ref{49}), (\ref{50}). Combinations of constraints in
Eq.(\ref{47}) are chosen in such a way that all mixed brackets
(i.e. those among Eq.(\ref{47}) and Eqs.(\ref{48})-(\ref{50})) vanish.
The fermionic constraints $L_\alpha$ obey the Poisson bracket algebra
(\ref{3}). The constraints (\ref{46}), (\ref{47}) are
first class. 

To proceed further, let us make partial fixation of gauge. One imposes
$N=1, ~ N_1=0$ for Eq.(\ref{46}) and $D^\mu\Gamma^\mu\chi=0$ for
1CC $\Lambda^\mu\Gamma^\mu p_\chi=0$ contained in Eq.(\ref{50}). After
that, the pairs $(N, p_N), (N_1, p_{N1}), (\chi, p_\chi)$ can be omitted
from consideration. The Dirac bracket for the remaining variables
coincides with the Poisson one. In the same fashion, the pair
$\eta, p_\eta$ can be omitted also. Then the remaining constraints (as
well as equations of motion) coincide with those of the GS superstring,
which proves equivalence of the actions (\ref{34}) and (\ref{44}).

On other hand, retaining the variables $\eta, p_\eta$ and the 
corresponding constraints (\ref{49}), the system
(\ref{48}), (\ref{49}) can be rewritten equivalently as in 
(\ref{11})-(\ref{13}), 
with the irreducible 1CC (\ref{11}), which are separated from the
2CC (\ref{12}), (\ref{13}).

Covariant and irreducible gauge for Eq. (\ref{11}) can be chosen as
\begin{eqnarray}\label{58}
R_\alpha\equiv\Lambda^\mu(\Gamma^\mu\theta)_\alpha+
D^\mu(\Gamma^\mu\eta)_\alpha=0,
\end{eqnarray}
or, equivalently
\begin{eqnarray}\label{59}
\Lambda^\mu(\Gamma^\mu\theta)_\alpha=0, \qquad
D^\mu(\Gamma^\mu\eta)_\alpha=0.
\end{eqnarray}
Matrix of the Poisson brackets
\begin{eqnarray}\label{60}
\{\Phi^{(1)\alpha}, R_\beta\}=[2(D\Lambda)\delta^\alpha_\beta+
4iTD^\mu(\tilde\Gamma^\mu\Gamma^\nu\partial_1\theta)^\alpha
(\Gamma^\nu\eta)_\beta]\delta(\sigma-\sigma '),
\end{eqnarray}
has a body on its diagonal and is invertible. It means that
Eqs.(\ref{11}), (\ref{58}) allows one to construct the Dirac bracket
without fermionic inconsistencies.

To conclude this section, let us discuss dynamics of the physical sector 
variables in the covariant gauge for $\kappa$-symmetry. 
Crucial property of the standard
noncovariant gauge $\Gamma^+\theta=0$ is that equations of motion in
this case acquire linear form. Then it is possible to find their
general solution. While not necessary for construction of the formal
path integral, namely this fact allows one to fulfill really the
canonical quantization procedure. Similarly to this, the covariant
gauge will be reasonable only if it has the same property.
To study equations of motion for physical variables, let us consider
the gauge $D^\mu\Gamma^\mu\chi=0$ and Eq.(\ref{59}). Then the variables
$(\chi, p_\chi), (\eta, p_\eta)$ can be omitted. 
The remaining constraints are (\ref{46})-(\ref{48}), which are 
accompanied by the covariant gauge condition 
\begin{eqnarray}\label{61}
\Lambda^\mu\Gamma^\mu\theta=0.
\end{eqnarray}
Equations of motion for the theory look now as follows
\begin{eqnarray}\label{65}
\partial_0x^\mu=-\frac{N}{T}p^\mu-N_1\partial_1x^\mu+
2iN\theta\Gamma^\mu\tilde P_-\partial_1\theta, \cr
\partial_0p^\mu=\partial_1[-TN\partial_1x^\mu-N_1p^\mu+
2iNT\theta\Gamma^\mu\tilde P_-\partial_1\theta], \cr
\partial_0\theta^\alpha=-N_1\partial_1\theta^\alpha-
\tilde K^\alpha{}_\beta\partial_1\theta^\beta,
\end{eqnarray}
where $\tilde P^\alpha_{\pm\beta}$ are covariant projectors on 
eight-dimensional subspaces [34]. In the standard gauge 
\begin{eqnarray}\label{23}
N=1, \qquad N_1=0,
\end{eqnarray}
for the constraints (\ref{46}) the equations of motion 
remain non linear. 
Nevertheless, usual picture of the Fock space can be obtained in the
covariant gauge. To resolve the problem one can use a trick which
we refer here as ``off-diagonal gauge'' for $d=2$ fields. 
Namely, let us consider
\begin{eqnarray}\label{29}
N=0, \qquad N_1=-1,
\end{eqnarray}
instead of (\ref{23}). Then Eq.(\ref{65}) acquires the linear 
form\footnote{While the action
(\ref{44}) is not well defined for the value $N=0$, the Hamiltonian
formulation (\ref{45})-(\ref{50}), (\ref{65}) 
admits formally Eq.(\ref{29}) as
the gauge fixing conditions for the constraints (\ref{46}).}
\begin{eqnarray}\label{66}
\partial_-x^\mu=0, \qquad \partial_-p^\mu=0, \qquad 
\partial_-\theta_a=0,
\end{eqnarray}
where $\theta_a, ~ a=1, \ldots ,8$ is $8_c$ part of $\theta^\alpha$,
while $8_s$ part $\theta_{\dot a}$ is determined by the covariant gauge
condition $\Lambda^\mu\Gamma^\mu\theta=0$. 

Fermionic dynamics is the same as in the light-cone gauge.  
Let us demonstrate that bosonic sector of (\ref{66}) leads also to the same 
description of state space as those of string in the usual gauge 
(\ref{23}). Actually,
solution of Eq.(\ref{66}) is ($0\le\sigma\le\pi$, closed string)
\begin{eqnarray}\label{30}
x^\mu(\tau, \sigma)=X^\mu+\frac{i}{\sqrt{\pi T}}\sum_{n\ne 0}\frac 1n
\beta_n^\mu e^{2in(\tau+\sigma)}, \cr
p^\mu(\tau, \sigma)=
\frac{1}{\pi}P^\mu-2\sqrt{\frac{T}{\pi}}\sum_{n\ne 0}
\gamma_n^\mu e^{2in(\tau+\sigma)}.
\end{eqnarray}
From these expressions one extracts the Poisson brackets for
coefficients. For the variables
\begin{eqnarray}\label{31}
\bar\alpha_n^\mu\equiv\beta_n^\mu+\gamma_n^\mu, \quad
\alpha_n^\mu\equiv\beta_{-n}^\mu-\gamma_{-n}^\mu, \quad
\alpha_0^\mu=-\bar\alpha_0^\mu=\frac{1}{2\sqrt{\pi T}}P^\mu,
\end{eqnarray}
one obtains the properties
\begin{eqnarray}\label{32}
\{\alpha_n^\mu, \alpha_k^\nu\}=
\{\bar\alpha_n^\mu, \bar\alpha_k^\nu\}=
in\eta^{\mu\nu}\delta_{n+k,0}, \quad
\{X^\mu, P^\nu\}=\eta^{\mu\nu}, \cr
(\alpha_n^\mu)^*= \alpha_{-n}^\mu, \qquad
(\bar\alpha_n^\mu)^*=\bar \alpha_{-n}^\mu.
\end{eqnarray}
In terms of these variables bosonic part of the Virasoro 
constraints $D^2=\Lambda^2=0$ acquire the standard form
\begin{eqnarray}\label{33}
L_n=\frac 12\sum_{\forall k}\alpha_{n-k}\alpha_k=0, \quad
\bar L_n=\frac 12\sum_{\forall k}\bar\alpha_{n-k}
\bar\alpha_k=0.
\end{eqnarray}
Eqs.(\ref{32}), (\ref{33}) have the same form as those of string in the
gauge (\ref{23}) and contain all the necessary information for 
canonical quantization [31-33]. Thus, instead of the standard gauge one
can equivalently use the conditions (\ref{29}) and the covariant gauge 
for $\kappa$-symmetry (\ref{61}), which gives the same
structure of state space. 

From Eq.(\ref{30}) it follows that $x^\mu(\tau, \sigma)$ is not true
string coordinate. If it is necessary, the latter can be restored
by using of the following statement (see [34]): \\
Let $\tilde x^\mu, ~ \tilde p^\mu$ represent general solution of
the system $\partial_-\tilde x^\mu=0$,
$\partial_-\tilde p^\mu=0$.
Then the quantities
\begin{eqnarray}\label{28}
x^\mu(\tau, \sigma)=\tilde x^\mu(\sigma^+\mapsto\sigma^-)-
\int_{0}^{\sigma^+}dl\tilde p^\mu(l), \cr
p^\mu(\tau, \sigma)=\frac 12[\tilde p^\mu-\partial_-\tilde x^\mu
(\sigma^+\mapsto\sigma^-)],
\end{eqnarray}
give general solution of the string equations of motion in the standard
gauge (\ref{23}) $\partial_0x^\mu=-p^\mu$,
$\partial_0p^\mu=-\partial_1\partial_1x^\mu$.

\section{Type IIB Green-Schwarz superstring with irreducible first 
and second class constraints.}

Here we show that type IIB theory is essentially different from other 
models with $\kappa$-symmetry. Namely, in this case {\em{both}} first and 
second class constraints can be arranged into irreducible sets 
{\em{in the initial formulation}}, without introducing of an additional 
variables. 

Denoting two Majorana-Weyl spinors of the same chirality as  
$\theta^{1\alpha}, ~ \theta^{2\alpha}, ~ \alpha=1, \ldots ,16$, the  
type IIB GS superstring action is
\begin{eqnarray}\label{70}
S=T\int d^2\sigma
\left[\frac{1}{2\sqrt{-g}}g^{ab}
\Pi_a^\mu \Pi_b^\mu-i\varepsilon^{ab}\partial_ax^\mu
(\theta^1\Gamma^\mu\partial_b\theta^1-
\theta^2\Gamma^\mu\partial_b\theta^2)+\right. \cr 
\left. \varepsilon^{ab}(\theta^1\Gamma^\mu\partial_a\theta^1)
(\theta^2\Gamma^\mu\partial_b\theta^2)\right],
\end{eqnarray}
where $\Pi^\mu_a\equiv
\partial_ax^\mu-i\theta^A\Gamma^\mu\partial_a\theta^A$. Let us denote
\begin{eqnarray}\label{72}
B_1^\mu\equiv p^\mu+T\partial_1x^\mu-iT\theta^1\Gamma^\mu
\partial_1\theta^1, \cr
B_2^\mu\equiv p^\mu-T\partial_1x^\mu+iT\theta^2\Gamma^\mu
\partial_1\theta^2, \cr
\hat p^\mu\equiv p^\mu-iT\theta^1\Gamma^\mu\partial_1\theta^1+
iT\theta^2\Gamma^\mu\partial_1\theta^2,  
\end{eqnarray}
\begin{eqnarray}\label{73}
D^\mu\equiv\hat p^\mu+T\Pi_1^\mu=
p^\mu+T\partial_1x^\mu-2iT\theta^1\Gamma^\mu\partial_1\theta^1, \cr 
\Lambda^\mu\equiv\hat p^\mu-T\Pi_1^\mu=p^\mu-T\partial_1x^\mu+
2iT\theta^2\Gamma^\mu\partial_1\theta^2,
\end{eqnarray}
Then constraints under the interest for Eq.(\ref{70}) can be written as 
\begin{eqnarray}\label{75}
H_+\equiv D^2-4TL^1_\beta\partial_1\theta^{1\beta}=0, \quad
H_-\equiv \Lambda^2+4TL^2_\beta\partial_1\theta^{2\beta}=0,
\end{eqnarray}
\begin{eqnarray}\label{76}
L^1_\alpha\equiv p_{\theta 1\alpha}-iB_1^\mu(\Gamma^\mu
\theta^1)_\alpha=0, \cr 
L^2_\alpha\equiv p_{\theta 2\alpha}-iB_2^\mu(\Gamma^\mu
\theta^2)_\alpha=0,
\end{eqnarray}
Poisson brackets for the fermionic constraints are 
\begin{eqnarray}\label{77}
\{L^1_\alpha, L^1_\beta\}=2iD_\mu{\Gamma^\mu}_{\alpha\beta}
\delta(\sigma-\sigma '), \cr 
\{L^2_\alpha,L^2_\beta\}=2i\Lambda_\mu{\Gamma^\mu}_{\alpha\beta}
\delta(\sigma-\sigma ').
\end{eqnarray}
As for $N=1$ case, the constraints $H_{\pm}$ were chosen such 
that they commutes with (\ref{76}). In contrast to $N=1$ case, one 
has now the fermionic constraints in duplicate form. On this reason 
the second step of supplementation scheme is {\em{not necessary}} 
here. Eq.(\ref{76}) can be rewritten equivalently as 
\begin{eqnarray}\label{78}
D^\mu(\tilde\Gamma^\mu L^1)^\alpha=0, \qquad 
\Lambda^\mu(\tilde\Gamma^\mu L^2)^\alpha=0;
\end{eqnarray}
\begin{eqnarray}\label{79}
\Lambda^\mu(\tilde\Gamma^\mu L^1)^\alpha=0, \qquad
D^\mu(\tilde\Gamma^\mu L^2)^\alpha=0, \qquad
\end{eqnarray}
where Eq.(\ref{78}) (Eq.(\ref{79})) contains 1CC (2CC) correspondingly. 
According to the statement 2), the equivalent system is 
\begin{eqnarray}\label{80}
\Phi^{(1)\alpha}\equiv
D^\mu(\tilde\Gamma^\mu L^1)^\alpha+
\Lambda^\mu(\tilde\Gamma^\mu L^2)^\alpha=0 \cr 
\Phi^{(2)\alpha}\equiv
\Lambda^\mu(\tilde\Gamma^\mu L^1)^\alpha+
D^\mu(\tilde\Gamma^\mu L^2)^\alpha=0.
\end{eqnarray}
Here $\Phi^{(1)\alpha}$ consist of 16 irreducible 1CC and 
$\Phi^{(2)\alpha}$ represents 16 irreducible 2CC. Non zero Poisson 
bracket is 
\begin{eqnarray}\label{81}
\{\Phi^{(2)\alpha}, \Phi^{(2)\beta}\}=-4i(D\Lambda)
(D^\mu+\Lambda^\mu)\tilde\Gamma^{\mu\alpha\beta}
\delta(\sigma-\sigma '),
\end{eqnarray}
and is manifestly non degenerated. 

Covariant and irreducible gauge for 1CC $\Phi^{(1)\alpha}=0$ 
can be chosen as 
\begin{eqnarray}\label{82}
R_\alpha\equiv\Lambda^\mu(\Gamma^\mu\theta^1)_\alpha+
D^\mu(\Gamma^\mu\theta^2)_\alpha=0.
\end{eqnarray}
The corresponding bracket 
\begin{eqnarray}\label{83} 
\{\Phi^{(1)\alpha}, R_\beta\}=[2(D\Lambda)\delta^\alpha_\beta+ 
4iT(D^\mu(\tilde\Gamma^\mu\Gamma^\nu\partial_1\theta^1)^\alpha
(\Gamma^\nu\theta^2)_\beta- \cr
\Lambda^\mu(\tilde\Gamma^\mu\Gamma^\nu\partial_1\theta^2)^\alpha
(\Gamma^\nu\theta^1)_\beta)]\delta(\sigma-\sigma '),
\end{eqnarray}
is manifestly non degenerated also.

\section{Type IIA Green-Schwarz superstring with irreducible second 
class constraints.}

The type IIA formulation can be obtained from Eqs.(\ref{70})-
(\ref{75}) by substitution $\theta^{2\alpha}\mapsto\theta^2_\alpha$ 
as well as $\Gamma^\mu_{\alpha\beta}\mapsto
\tilde\Gamma^{\mu\alpha\beta}$ in each place where $\Gamma$-matrix 
is associated with the spinor $\theta^2$. The fermionic constraints are 
\begin{eqnarray}\label{84}
L^1_\alpha\equiv p_{\theta 1\alpha}-iB_1^\mu
(\Gamma^\mu\theta^1)_\alpha=0, \cr
L^{2\alpha}\equiv p_{\theta 2}^\alpha-iB_2^\mu
(\tilde\Gamma^\mu\theta^2)^\alpha=0. 
\end{eqnarray}
They belong to two inequivalent representations of $SO(1, 9)$ group 
and thus can not be recombined in the Poincare covariant way. 
Moreover, there is no of natural object in the problem which can be 
used for raising (lowering) of spinor index. Nevertheless, after 
separation of the constraints, part of them can be arranged into 
irreducible subsets. Repeating the same procedure as in $N=1$ case, 
one obtains the additional sector 
\begin{eqnarray}\label{85}
\Lambda^\mu(\Gamma^\mu\eta)_\alpha=0, \qquad p_{\eta\alpha}=0.
\end{eqnarray}
Then Eqs.(\ref{84}), (\ref{85}) can be rewritten in the equivalent form 
\begin{eqnarray}\label{86}
\Phi^{(1)\alpha}\equiv
D^\mu(\tilde\Gamma^\mu L^1)^\alpha+
\Lambda^\mu(\tilde\Gamma^\mu p_\eta)^\alpha=0,
\end{eqnarray}
\begin{eqnarray}\label{87}
\Lambda^\mu(\Gamma^\mu L^2)_\alpha=0,
\end{eqnarray}
\begin{eqnarray}\label{88}
\Phi^{(2)\alpha}\equiv
\Lambda^\mu(\tilde\Gamma^\mu L^1)^\alpha+
D^\mu(\tilde\Gamma^\mu p_\eta)^\alpha=0, \cr
G^{(2)\alpha}\equiv
D^\mu(\Gamma^\mu L^2)_\alpha+
\Lambda^\mu(\Gamma^\mu\eta)_\alpha=0,
\end{eqnarray}
Here $\Phi^{(1)\alpha}$ are irreducible 1CC while 
$\Phi^{(2)\alpha}\, ~ G^{(2)\alpha}$ form irreducible set of 2CC. Thus, 
only 1CC in Eq.(\ref{87}) remain reducible.

\section{Conclusion}

In this work we have analysed a possibility to apply the 
supplementation scheme [26] to the Green-Schwarz superstring. It was 
shown that type IIB theory represents an exceptional case, namely, 
both first and second class constraints can be arranged into irreducible 
sets (\ref{80}) in the initial formulation. It implies that the standard 
path integral quantization methods can be applied to the resulting 
system (\ref{80}) in a manifestly covariant form. 

For the type IIA theory only 2CC can be arranged into irreducible 
set (\ref{88}).

For both cases consistent Dirac bracket can be constructed and then the 
corresponding 2CC can be eliminated in a covariant way. It leads to 
the intriguing suggestion that it may exist manifestly $D10$ 
supersymmetric action which is equivalent to Type II GS superstring and 
which contains 1CC only. 

For $N=1$ case we have proposed the modified action (\ref{44}). 
In addition to usual superspace coordinates it involves a 
pair of the Majorana-Weyl spinors. The additional variables are subject 
to reducible constraints (\ref{49}), (\ref{50}), which supply their 
nonphysical character (see discussion after Eq.(\ref{10})). Equivalence 
of the modified action and the initial one was proved in the canonical 
quantization framework (see discussion after Eq.(\ref{50})). We have 
demonstrated also how the state spectrum can be studied in the covariant 
gauge for $\kappa$-symmetry.
In the modified formulation first class constraints form irreducible set 
(\ref{11}) and are separated from the second class one (\ref{12}), 
(\ref{13}). The 
corresponding covariant gauge (\ref{58}) is irreducible also, which 
garantees applicability of the usual path integral methods for the 1CC 
sector of the theory. 

\section*{Acknowledgments.}

The work has been supported partially by Project GRACENAS No 97-6.2-34.

\end{document}